\documentstyle[12pt]{article}
\newcommand{\be}{ \begin{eqnarray}}
\newcommand{\ee}{\end{eqnarray}}
\newcommand{\beno}{ \begin{eqnarray*}}
\newcommand{ \eeno}{\end{eqnarray*}}
\newcommand{\bfg}{\begin{figure}[h]}
\newcommand{\efg}{\end{figure}}
\newcommand{\raf}[1]{(\ref{#1})}

\newcommand{\ch}{{\chi S R}}
\begin{document}
\bibliographystyle{unsrt}
\setlength{\baselineskip}{24.2pt}
\begin{titlepage}
\begin{flushright}
SUNY-NTG-92-4
\end{flushright}
\vspace{1cm}
\begin{center}
{\bf \large The Propagation of Quarks in the
Spatial Direction in Hot QCD}
\footnote{Supported by the U.S. Dept. of Energy
Grant No. DE-FG02-88ER40388}\\
\vspace{1.5cm}
\ \\
\ \\
{\large V. Koch\footnote{Supported by Deutsche Forschungsgemeinschaft DFG},
E.V. Shuryak, G.E. Brown, and A. D. Jackson}
\ \\
Department of Physics\\
State University of New York at Stony Brook\\
Stony Brook, NY 11794 \\
\ \\
\vspace{1.5cm}
\newpage
{\large \bf Abstract}\\
\end{center}
The dynamics of {\it light} fermions propagating in a spatial direction
at high temperatures can be described effectively by a two--dimensional
Schr\"odinger equation with {\it heavy}
effective mass $m_{\rm eff} = \pi T$.
Starting from QED, we discuss
the
transition from three-- to two--dimensional positronium discussing the latter
 in detail including relativistic effects.
In the case of QCD the problem is similar to that of heavy quarkonium.
Our effective potential contains the usual Coulomb and confining
parts as well as a perturbative spin--spin interaction. The resulting
$\bar q q$ ``wave functions" reproduce
recent lattice data for the $\rho$ and $\pi$ channels.
The physical meaning of such `confinement' is
related to the non--trivial magnetic interaction of color currents
in the quark--gluon plasma.  Our results  do not contradict
the idea that the normal
electric interaction
of color charges is screened and produces no bound states in the usual sense.
\vspace{0.2cm}

\end{titlepage}
 \newpage
\centerline{\bf 1. Introduction}
\ \\
It is generally believed that hadronic matter at  high temperature
undergoes a transition to a new phase, the so called {\it quark-gluon plasma}
(QGP), which is a gas of weakly interacting quarks and gluons. As far as
global properties ({\em e.g.,} thermodynamic properties) are concerned,
this statement is supported by both perturbative QCD analyses and
numerical simulations based on Lattice Gauge Theory (LGT). (See, {\em e.g.,}
general reviews \cite{GROSS_PISARSKI_YAFFE_1981}, \cite{Shuryak_80}).)
However, this does not imply the absence of non--trivial phenomena
even at very high $T$. As an example, one may mention the problem of
magnetic screening, the mechanism for which remains unknown.

Some time ago, DeTar \cite{DeT85,DeT88} suggested that the QGP may be
`dynamically  confined', in the sense that only color--singlet modes
produce poles and branch points in linear response functions. In other words,
he proposed that color--singlet modes might control the
large--distance behavior of plasma disturbances. DeTar and Kogut
\cite{DK87a,DK87b} pursued this matter numerically by calculating
the correlation functions of various operators using LGT. The `technical'
point of importance for the interpretation of these results is that
the correlations were measured in the {\em spatial} direction.

These functions were found to decay exponentially with distance,
$\sim \exp(-M z)$, and DeTar and Kogut \cite{DK87a,DK87b}
studied the temperature dependence of the exponent, $M$, (known
as the screening mass) in a number of channels. The
screening masses of the chiral partners
($\pi, \, \sigma$), ($\rho,  \, A_1$) and ($N(\frac12 +), \, N(\frac12 -)$)
were found to become equal within the
accuracy of the calculations at the temperature
where chiral symmetry is restored
(defined as the point where $<\bar q q> \rightarrow 0$ and referred to
below  as $T_c = T_\ch$).

For spin zero channels, ($\pi$,$\sigma$), and in the chiral limit, one would
expect the mass to remain small even above $T_c$.  These modes can be viewed
as long--range fluctuations in
the order parameter of the {\it second--order} chiral restoration transition
as first proposed by
 Hatsuda and Kunihiro \cite{HK84,HK85,HK85b}.  Indeed, lattice data
show that $M_\pi(=M_\sigma)$
remains relatively small  until rather large $T$.  (Further discussion
of this point and a compilation of data can be found in \cite{Gocksch}.)
In contrast to this result,
Eletskii and Ioffe \cite{EI88} noted that for the other
doublets the following is
true:
 $M_\rho = M_{A_1} \simeq 2 \pi T=2\omega_0$.  For the baryonic
doublet the mass is about $3 \pi T=3\omega_0$.  These values are essentially
those suggested by the lowest quark Matsubara frequency, $\omega_0=\pi
T$, and led to the suggestion that these correlators
can be interpreted as describing the {\it independent} motion of two (or three)
quarks in a plasma described as an ideal quark gas.

For people not comfortable with the Matsubara formalism, one can
explain this point in a different way by looking at equal time correlations,
as recently suggested by one of us
\cite{Shuryak_talks}.
One can show that the  propagator
for  free massless quarks at temperature, $T$, over a
{\it spatial} distance $x$ in {\it zero} time is given by
$$
S_T(x)=(\gamma_x \partial_x) \int {d^3k \over (2\pi)^3 k}
\exp(i{\vec k} \cdot {\vec x})[{1 \over 2} -
{1 \over 1+\exp(k/T)}] .
$$
The last term contains Fermi occupation numbers, and the combination
($1/2-n_F(k)$) is clearly related to the familiar combination ($1/2+n_B(k)$)
in the case of   Bose statistics. In both cases the `1/2' term is  due to
zero--point oscillations and the last term to the contribution of
thermal excitations.  However, quarks with small momentum
have $n_f=1/2$.  Thus, one has a cancelation of these two contributions
for $k \ll T$ which eliminates power contributions and leads to
$S_T(z) \sim \exp(-\pi T z)$.  Another amusing way to look at this comes by
rewriting $(1/2-n_F(k))$ as $(1/2)[(1-n_F)-n_F]$ so that the particle--hole
symmetry $n_F \leftrightarrow (1-n_F)$ is made obvious. Since the situations
near $n_f=0$ and $n_f=1$ are physically similar (up to the sign of the
propagator), it is natural to expect that the midpoint, $n_F=1/2$, is special.
There, the particle and hole terms cancel.

It is useful to mention that there are essentially two simple
ways to make the connection between the imaginary--time formalism and
real time. One is to look at equal time correlations ({\em i.e.,} correlators
integrated over frequency) as in the previous illustration. The other
possibility is to study correlators at zero frequency ({\em i.e.,}
correlators integrated over imaginary time.)  It is the latter which is used
to extract the screening mass in LGT calculations while the former is used to
determine the wave functions.

The scenario of free motion in a ideal quark gas is too simplistic
to describe the real situation, but the observation that most screening
masses are consistent with this picture suggests that other mechanisms have
little net effect.  In particular, these data provide an important limitation
on any dynamically generated quark mass, $m_{dyn}$ \cite{BBJ91}.  The
screening mass for massive quarks would be $M = \sqrt{\omega_0^2 + m_{dyn}^2}$.
Thus, it seems that an appreciable $m_{dyn}$ is excluded even in the region
$T \sim T_\ch$.  A picture in which hadrons go massless \cite{BR91} as
$T \rightarrow T_\ch$ (leading to the scenario suggested in \cite{BBP91,BBJ91})
is not excluded.  (However, given a temperature of $T \simeq 150\, \rm MeV$, an
accuracy of about 5\% for the calculated screening masses is required
to rule out dynamical quark masses smaller than $m_{dyn} = 150$ MeV.)

The next important step was made by recent lattice studies \cite{BDD91} which
revealed that the $\bar q q$ pair, moving over large distances in the spatial
direction, is actually closely correlated in the transverse plane.
Moreover, this correlation is even more pronounced at finite $T \simeq 150$
MeV than it is at $T=0$. It was suggested \cite{BDD91} that, in line with
DeTar's dynamical confinement ideas, the quark and antiquark are not free at
all but bound even at high $T$. The aim of the present paper is to provide a
quantitative explanation of these data as well as a general discussion of
what these data actually mean for the physics of the QGP.

An essential technical point, important for the understanding of the
terminology used, is the interchange of the time and $z$--axis which we shall
describe in some detail in the next section.  After  transformation to this
`funny space',  one deals with a system at {\em zero} temperature but
placed in a box periodic in the $z$--direction with periodicity $\beta=1/T$.
Using this language it is obvious that, in the high temperature limit, a
so--called dimensional reduction takes place.  The $(3+1)$--dimensional
gauge theory becomes a $(2+1)$--dimensional theory as suggested long
ago \cite{AC75}.  The main physical point is that, in the high $T$ limit,
the motion of a quark in the `funny space' is dominated by its momentum in
the $z$--direction which is governed by $\pi T$ as a consequence of the
antiperiodic boundary conditions.  For motion in the transverse direction, as
we shall show, this momentum behaves like a mass.  Since at high temperature
this `mass' becomes very large, {\em any} attractive potential can bind
quarks in the transverse direction.

The main objective of this work is to derive the high--$T$ limit of the
``wave functions'' introduced (and measured on the lattice) in  \cite{BDD91}.
In section 2 we start with the QED case and discuss how the transition
from $d=3$ to $d=2$ positronium takes place. Qualitative features of
$d=2$ bound states in the $d=2$ logarithmic Coulomb potential were recently
discussed in ref.\cite{HZ91}. We report the results of a more quantitative
analysis including next--order corrections. We pay special attention to
the spin--spin interaction because this issue is important for the
mesonic splittings discussed in section 3.

The effective $d=2$ QCD potential for a $\bar q q$--pair
propagating in the spatial direction was studied  on the lattice in
ref.\cite{MP87}.  It was found to be essentially temperature independent.
(This result stands in sharp contrast to the usual potential, for propagation
in time, which exhibits both deconfinement and charge screening.)  Although
the data \cite{MP87} are  not very good at large separations and allow a wide
range of interpretations, comparison with the ``wave functions'' allows us
to conclude that the ``string tension'' is essentially $T$--independent
in the interval between $T=0$ and $T = 210$ MeV studied in ref.\cite{BDD91}.
Another issue addressed in section 3 deals with the differences, both
in the shape of the ``wave functions'' and the screening masses, between the
$\pi$ and $\rho$ systems.  A spin--spin interaction of the
Fermi--Breit form provides a natural explanation of most of these differences.

In section 4 we proceed to a discussion of the physical meaning of these
``bound states''.  In order to understand these results, one must come back
from the `funny space' to ordinary space.  One realizes that by
dealing with static correlation functions (with no time involved), it is not
possible to say anything about the energy spectrum of excitations in the QGP
at high temperatures.  What is actually studied is the momentum spectrum
of the correlators for various channels.  At $T=0$ the two spectra are
identical.  At high $T$, they are completely different.
\ \\

\newpage

\centerline{\bf 2.  Propagating an $e^+ e^-$ pair at high T}
\ \\
In this section we shall study the ``wave function'' of an $e^+ e^-$
pair at finite temperature in Euclidian space. We shall proceed by ignoring the
Coulomb force and restricting our attention to the magnetic interaction
between currents.  The aim of this restriction is to preserve the analogy to
QCD in which the electric interaction is screened and magnetic interactions
are believed to be dominant at high temperature.  The purpose of this example
is that QED is an abelian gauge theory with a fixed and small coupling
constant, $e$.  One can work out all essential ingredients, and both the
``binding energy'' and the ``wave function'' of an $e^+ e^-$ pair at high $T$
can be obtained.  As is well known, a quantum field theory at finite
temperature, $T$, can be formulated  using Euclidean time, $\tau$, ranging
from $0$ to $\beta = 1/T$ with  periodic boundary conditions for bosons and
antiperiodic boundary conditions for fermions. This formulation, which is
the corner stone of LGT calculations, will be the starting point
for our considerations.  This convenient generalization of the $T=0$
case has, however, its limitations.  One is unable to consider any
time--dependent quantities and deals only with static (heat bath averaged)
quantities.  As a consequence, LGT cannot tell us anything about the
frequency spectrum of elementary excitations.  What we are going
to study in this work is, therefore, related to the momentum spectrum.  (In
other words, properties related to {\it spatial} correlation functions.)

Given the magnetic nature of this problem, it is simpler to proceed by
interchanging the (compressed) Euclidean time, $\tau$, and the spatial
direction, $z$, in which the correlator is measured.  This transformation leads
us to a `funny space' where the system propagates in unlimited time (hence, at
zero temperature) but is confined in a  box in the $z$--direction.  This
transformation must be applied to the fields.  Specifically, the original
magnetic field, $F_{13}$ and $F_{23}$, becomes an electric field ({\em i.e.,}
$F_{13} \Rightarrow F_{10} = E_x$ and $F_{23} \Rightarrow F_{20} = E_y$).
The magnetic interaction between two rapidly moving charges is
transformed into an electric interaction between two, nearly static, charges.
We emphasize that this new Coulomb interaction has emerged from a technical
manipulation.  Its physical origin remains magnetic and, as such, it is
not subject to the effects of screening which are presumed to suppress
genuine electric interactions.

The presence of a spatially periodic ``box'' modifies the interaction of the
charges.  This modification is readily calculated for our simple example
of small ($\alpha=e^2 \ll 1$) charges in QED with dynamics given by the
current--current interaction of the leading one--photon exchange diagram.
The $\tau z$--interchange immediately transforms this interaction into
the ordinary Coulomb potential, $V=-\alpha/R$.  The presence of the ``box''
boundaries modifies this result significantly because the electric potential
must now satisfy periodic boundary conditions in the $z$--direction.  There
are two ways of solving the standard Poisson equation:
\be
- \nabla^2 \Phi = \rho .
\label{pois}
\ee
One can either (i) expand the field  in Fourier components
\be
\Phi(x,y,z) = \sum_n \Phi_n(x,y) e^{-i 2 n \pi T \, z}
\nonumber \\
\rho(x,y,z) = \sum_n \rho_n(x,y) e^{-i 2 n \pi T \, z}
\label{eq:2.2}
\ee
(in the high T limit only the n=0 modes are important) or (ii) one can
construct a periodic array of ``reflected charges".  (See Figure 1.)
The second approach is certainly simpler; it involves no integrals and leads
directly to the sum
\be
\Phi = {e \over 4\pi }[{1 \over r} - 2 \sum_{n=1}^\infty
({1 \over \sqrt{ x^2 +y^2
+(n\beta-z)^2}}-{1 \over n\beta})].
\label{eq:2.3}
\ee
We note that the sum is automatically regularized at large $n$ and that the
large--$T$ limit of this three--dimensional problem is simply the
two--dimensional Coulomb potential:
\be
\Phi_0(r_\bot) = -\frac{e}{2 \pi \beta} \ln(\frac{r_\bot}{2 \beta})
\label{eq:2.4}
\ee
which is equivalent to the potential of a uniformly
charged wire (with charge density $e/\beta$) in three dimensions.

The fermion field is also affected by the $\tau z$--interchange.  This is most
easily appreciated by considering the Dirac equation for a single, massless
particle in Euclidean space which has the form:
\be
\gamma_\mu^E \partial_\mu  \psi = 0
\ee
where the $\gamma_\mu^E$ are Euclidean gamma matrices.  The solutions to this
equation are
\be
\psi(x) = \psi_0 e^{-i (2n + 1) \pi T x_0} \, e^{-i \vec{p} \cdot \vec{x} }
\label{eq.1.2}
\ee
where the dependence on the $0$--direction is due to the antiperiodic boundary
conditions which result from the finite temperature.  (Here, $\psi_0$ is a
spinor which will not concern us.)  The Euclidean Dirac equation for this
problem is evidently invariant under the $\tau z$--interchange, and all effects
are concentrated on the wave function, eqn.(6), which becomes
\be
\psi(x) = \psi_0 e^{-i (2n + 1) \pi T z} \, e^{-i p_0 x_0} \,
e^{-i \vec{p}_\bot \cdot \vec{x}_\bot }
\ee
because of the boundary conditions.  The Dirac equation is solved by the choice
\be
p_0 = \pm i \, \sqrt{p_z^2 + p_\bot^2}  =
\pm i \, \sqrt{ ((2n+1) \pi T)^2 + p_\bot^2}  \equiv - i E
\ee
with $E$ real. The wave function thus takes the form
\be
\psi(x) = \psi_0 \, e^{ - E x_0}e^{-i (2n + 1) \pi T z}  \,
e^{-i \vec{p_\bot} \cdot \vec{x_\bot} }
\ee
which is the solution of a Dirac equation for a particle with a given momentum,
$p_z = (2n+1) \pi T$, in the $z$--direction.  For large $x_0$ only the state
with lowest energy, $E$, survives.  This state corresponds to the ground state
of the associated real time Dirac equation
\be
\gamma_0^M E \, \psi = ( \gamma_\bot^M p_\bot + \gamma_3^M (2n + 1) \pi T) \,
\psi
\ee
where the superscript,$M$, now indicates Minkowski $\gamma$-matrices.
This equation is equivalent to a Dirac equation in $(2+1)$--dimensions with a
``chiral'' mass of $m = (2n +1) \pi T$.  This can be rotated into a real mass
by
the unitary transformation
\be
\psi \rightarrow e^{- i \frac{\pi}{4} \gamma_3} \, \psi .
\ee
Thus, we have
\be
\gamma_0^M E \psi = ( \gamma_\bot^M p_\bot + (2n + 1) \pi T) \psi .
\ee
So far we have discussed the case of a massless particle.  For massive
fermions, one should use an effective mass of
\be
m_{\rm eff} = \sqrt{m_{dyn}^2 + [(2n +1) \pi T]^2}.
\ee
Since lattice gauge calculations provide no conclusive evidence of a dynamical
mass, we will restrict ourselves to the case of vanishing dynamical mass.

It is clear from eqn.(13) that the effective mass of the electron is very
large at high temperature.  The dynamics thus become nonrelativistic and
can be described by a Schr\"odinger equation in two dimensions:
\be
E \, \psi = \frac{p_\bot^2}{2 m} \, \psi  .
\label{schr}
\ee
The effects of any additional potential ({\em e.g.,} an ``electrostatic''
potential, $V = -e \Phi$) can then simply be included by making the
substitution $E \rightarrow E-V$ in eqn.\raf{schr}.

These manipulations are readily extended to the present example of an
interacting $e^+ e^-$ pair. (The details are given in the Appendix.)
We note that the restriction to zero Matsubara frequency, $n=0$, in
eqn.\raf{cpl} represents a decision and not an approximation.  It is also
desirable to make the approximation of neglecting all terms, $\psi^n_k$,
in eqn.\raf{cpl} with $k \not = 0$.  As we shall indicate shortly,
this approximation
introduces a tolerable relative error of ${\cal O}(e^2)$ for large $T$.

Before turning to numerical results, it is useful to perform a dimensional
analysis aimed at understanding the high $T$ behavior of this system.
For a wave function of size, $R$, the kinetic energy is equal to
$(\pi T R^2)^{-1}$.  Neglecting the binding energy, $E$,
we obtain
\be
\frac{1}{\pi T R^2} \simeq e^2 T .
\ee
Thus, the size of the bound system and the binding energy are
\be
R \sim \frac{1}{e T} \ \ \ {\rm and} \ \ \ E \sim e^2 T.
\ee
The size of the system is identical to electric screening length in
a hot plasma which is also $(eT)^{-1}$ . \cite{Shuryak_80}

With the approximations indicated above, the reduced mass for the $e^+ e^-$
pair becomes $m = \pi T / 2$, and the Schr\"odinger equation, eqn.\raf{cpl},
assumes the form
\be
-\frac{1}{\pi T} \left( \frac{1}{r} \frac{d}{dr} (r \frac{d \psi}{dr}) +
l^2 \psi \right) + V_{eff}(r) \psi = E \psi.
\label{eq.3.4}
\ee
Using dimensionless units,
\be
x = \frac{e r}{\beta}  ,
\ee
we obtain for the $l=0$ state
\be
-\left( \frac{1}{x} \frac{d}{dx} (x \frac{d \psi}{dx})
\right) + \tilde{V}_{eff}(x) \psi = \tilde{E} \psi
\label{eq.3.8}
\ee
where
\be
\tilde{E} = \frac{\pi E}{ e T } + 2 \pi \ln ( 2 T a)
\ee
and
\be
\tilde{V}_{eff}(x) = 2 \pi \, \ln (x).
\label{eq.3.10}
\ee
Note that eqn.\raf{eq.3.8} no longer depends on the physical parameters.
Here, we have used the high temperature limit of the effective
potential given by eqn.\raf{eq:2.4}.  The wave function is now given in terms
of the dimensionless  variable, $x$, rather than the physical parameters,
$e$ and $T$. We have solved eqn.\raf{eq.3.8} numerically and the
resulting wave function is plotted together with the
potential in figure \ref{fig:pot}.  Restoring dimensions, the energy
is
\be
E= e^2 T (1.45 - 2 \ln(e)).
\ee

We now consider a variety of corrections to these high--$T$ results which, in
the approximation of zero dynamical mass, can be expressed purely in powers
of our dimensionless parameter, the electric charge $e$.  In particular, we
consider (i) the non--asymptotic part of the potential, (ii) excitation of
fermion modes with higher relative Matsubara frequencies ($k\not=1$), and (iii)
relativistic corrections.

In figure 3 we compare the modified Coulomb potential in the periodic box
given by eqn.\raf{eq:2.3} with its high--$T$, two--dimensional limit for a
few values of the temperature.  One can see that, although agreement is
generally good enough, there are significant deviations at small distances:
The original Coulomb potential, $e^2/4\pi r$, is
still present in the original form.
Since a typical separation is $r \sim 1/e T$, this produces corrections to
the energy of the system of ${\cal O}(e^3 T)$.

The approximation of neglecting all terms in eqn.\raf{cpl} with $k \not = 0$
can be
relaxed perturbatively.  Applying standard second--order perturbation theory
one obtains
\be
\delta E = - \sum_k {<1| V |k>^2 \over E_k-E_1} \sim e^4 T  .
\ee
Estimating the matrix elements as $V \sim {\cal O}(e^2T)$ and all
energy denominators as ${\cal O}(T)$, one finds this correction to be
$\delta E \sim e^4 T$.  As expected, this correction is smaller than the
leading contribution to the energy, eqn.(25), by a factor of ${\cal O}(e^2)$.
Taking only the first excited ($k=1$) state into account
numerical integration has produced the following
result for $e^2 = 0.2$ :
\be
\frac{\delta E}{E} = 3.5 \times  10^{-4}
\ee

A typical velocity, $v_t$, transverse to the compact $z$--axis is
$p_t/m_{\rm eff} \sim 1/RT \sim e$.  (The somewhat surprising feature of this
result is that it is not $e^2$ as it would be for the $d=3$ hydrogen atom.)
Therefore, relativistic corrections start
at the level $E_{\rm rel} \sim v_t^2 E_{\rm non-rel} \sim e^4 T$.  A
relativistic expansion similar to that leading to the Fermi--Breit Hamiltonian
can be performed as in the $d=3$ case.  We again have $-p^4/8m^2$ kinetic
energy corrections, the spin--orbit term (which does not affect
the ground state), spin--spin and ``Darwin'' terms.

We shall be concerned only with the properties of the spin--spin interaction
since we wish to discuss the spin splitting in QCD in some detail in the next
section.  In the $d=2$ space one has symmetry only with respect to rotation
in the $xy$--plane. Hence, states should be characterized by the
$z$--component of their spin. In the ground state (with vanishing
$z$--component of orbital angular momentum), the $e^+ e^-$ pair has two
degenerate states with total spin $S_z=\pm 1$ and two distinct
states with $S_z=0$ of different symmetry representing total spin $S=1$ and
$S=0$ (ortho and para-positronium, respectively).

We start with with the textbook ($d=3$) expression for the Fermi--Breit
interaction between two opposite charges
\be
V_{Breit}= {e^2 \over 4 m^2}[-{\vec \sigma_1 \cdot \vec \sigma_2 \over R^3}
+{3(\vec \sigma_1 \cdot \vec R)( \vec \sigma_2 \cdot \vec R)\over R^5}+{8\pi
\over 3} \vec \sigma_1 \cdot \vec \sigma_2 \delta(\vec R)]
\label{eq.25}
\ee
which consists of the familiar ``tensor force'' between magnetic moments and
a contact interaction. In addition, in the case of positronium, one has
to include an ``annihilation'' contribution\footnote{In the
next section (in which we discuss the quark--antiquark interaction), the
annihilation term is {\em not} present.}
\be
 V_{ann}= {\pi e^2 \over 2 m^2} (3+\vec \sigma_1 \cdot \vec \sigma_2)
 \delta(\vec R) .
\ee
Starting from these formulae, we can obtain the $d=2$ effective interaction
(needed in the high $T$ limit) by averaging these potentials over
$z$--direction. In three dimensions the tensor force vanishes for
the spherically symmetric ground state.  However, for finite temperature,
the ground state has a ``pancake'' shape and a non--zero
quadrupole moment.  This leads to a contribution from the tensor force.
This can be seen by rewriting eqn.\raf{eq.25} in momentum space as
\be
  V_{Breit} = {4 \pi e^2 \over 4 m^2}
\int {d^3q \over (2\pi)^3} \, e^{i\vec q \cdot \vec R} \,
[\vec \sigma_1 \cdot \vec \sigma_2 -
{(\vec \sigma_1 \cdot \vec q)( \vec \sigma_2 \cdot \vec q)\over q^2}] .
\ee
Averaging over $z$ means that we select the component of the potential which
has no momentum transfer in the $z$--direction. Therefore, the second term
contributes only if the spins are in the $xy$--plane:\footnote{It is useful to
consider an alternate derivation of the spin--spin interaction which
offers a different explanation for why spins in the $z$--direction represent
a special case.  As for the Coulomb field in the box,
one can satisfy periodic boundary conditions for the magnetic field
by using a periodic array of reflected magnetic moments and by then noting
that the $z$--directed moments display a strong tendency to cancel.
A nice way of explaining this was suggested to us by H.A.~Bethe.  The Coulomb
potential is a coherent sum of all reflections as they are all of the same
sign.  At high $T$, the potential of a charged wire results.  Magnetic
moments, on the other hand, can be considered as a pair of magnetic monopoles
sitting on this wire with alternating signs. From this it is clear that
the magnetic interaction vanishes since, for large $T$, the distance
between these monopoles decreases.}
\be
T\int dz \ V_{Breit}&=& {e^2T \over 4 m^2}
\int {d^2q_t \over (2\pi)^2} \ e^{i\vec q_t \cdot \vec R} \
[\vec \sigma_1 \cdot \vec \sigma_2 -
{(\vec \sigma_1 \cdot \vec q_t)( \vec \sigma_2 \cdot \vec q_t)\over q^2_t}]
\nonumber \\
&=&
{e^2T \over 4 m^2} \ 4\pi \ \delta^2(R_t) \ [\vec \sigma_1 \cdot \vec
\sigma_2-{1\over 2}
\vec \sigma_{1t} \cdot \vec \sigma_{2t}] .
\ee
\newpage
\noindent
The remaining local terms simply lead to two--dimensional $\delta$-functions.
\be
{1 \over\beta} \int_0^\beta dz \ \delta^3(\vec{r})=T \delta^2(\vec r_t) .
\ee
As a result, the spin splitting of $d=2$ positronium is
\be
E_{spin-spin}= {4\pi e^2 T |\psi(0)|^2\over  m^2}C_n
\ee
with the spin factors $C_n=3/4,1/2, -1/2$ for $S=1,S_z=\pm 1$,
$S=1,S_z=0$ and $S=0,S_z=0$ states, respectively.\\
\newpage
\centerline{\bf 3. $\bar q q$ ``binding'' in hot QCD }
\ \\
As noted in the introduction, there exists a significant general
difference between QCD and QED.  Even at arbitrarily high temperature, one is
confronted with non--trivial manifestations of the non--abelian nature of the
former theory as one can see from the ``dimensional reduction'' argument
\cite{AC75}.  The resulting $d=2+1$ theory is far from simple.  In
particular, as argued in \cite{dHoker,Borgs}, it is known to produce a linear
confining potential just as $d= 1+1$ and $d= 3+1$ (real QCD) theories do.
In QED the electron charge is a physical parameter, and this can be
{\it assumed} to be small in order to make the perturbative theory
meaningful.  In QCD we have a ``running'' coupling constant, and
the actual magnitude of $\alpha_s=g^2/4\pi$ depends on the distances (or
momenta) relevant for a given application.  The precise value can only be
specified if higher--loop corrections are calculated.  In the high--$T$
limit, for example, we have a large effective mass and therefore the size of
the Coulomb bound state is small, $R\sim {\cal O}(1/gT)$.\footnote{This is
similar to the case of superheavy quarkonium \cite{Appelquist_Politzer}
with the obvious substitution $e^2 \rightarrow (4/3)(g^2 / 4 \pi)$.}
However, the logarithmic corrections are actually cut off at the smaller
scale of $1/T$.  Therefore, (in contrast to quarkonium) one should expect
that the relevant $\alpha_s$ is $\alpha_s(T)$ which becomes small as
$T \rightarrow \infty$.

Unfortunately, the lattice data which we wish to discuss do not correspond to
this relatively clean limit because they have been obtained at temperatures
close to the critical temperature ({\em i.e.,} $T \simeq 1.5 T_c$ )
which corresponds
to effective masses of $m_{\rm eff} =\pi T \sim 450-600$ MeV.  From the
experience of $J/\psi$ and $\Upsilon$ physics, we know that these masses are
not sufficiently large to justify use of the Coulomb force alone.  However,
we have also learned another lesson from these examples:  A non--relativistic
approach based on a more complicated effective potential (which includes
confining forces) works well in describing quarkonium bound states.  It is
reasonable to try the same approach in our ``not--so--high'' $T$ case.
In practice this means that, as in the previous section, we will rotate into
the `funny space' and solve a Schr\"odinger equation for quarks of mass
$m_{\rm eff} = \pi T$.  Consequently, the effective potential must also be
given in this rotated space.  Fortunately, such an interaction has been
extracted on the lattice at high $T$ for pure glue \cite{MP87}.  The results
are shown in figure \ref{fig:potpol}.  Manosakis and Polonyi fit their results
using the standard parameterization
\be
V=- a/r+ \sigma r+ {\rm constant}
\label{eq:31}
\ee
with $a=0.184\pm 0.02$ and $\sqrt \sigma=0.22\pm 0.03 $.  These results can be
compared with the values $a=0.26$ and $\sqrt \sigma=0.22\pm 0.02$ obtained for
$T=0$.  If one identifies $\sigma$ with the phenomenological value of the
string tension, $(400 \ {\rm MeV})^2$, the physical
size of the lattice spacing is
$a \sim 1/(2 \ {\rm GeV}) \sim 0.1 \ {\rm fm}$ and $T \sim (1/3) \ {\rm GeV}$.
The constant in eqn.\raf{eq:31} was not evaluated in ref.\cite{MP87}.  It is
also instructive to compare these data with the potential appropriate for
quarks propagating in the time direction which is also shown in figure
\ref{fig:potpol}. Not only are the confining forces absent in this case
(because we are well above the deconfinement temperature), but the
Coulomb potential is much smaller due to screening effects.  In contrast to
these results, the potential measured in the  spatial direction
appears to be essentially $T$--independent, apart from some change at small
distances.  (This difference may simply result from the decrease of
the effective Coulomb charge due to a difference in radiative corrections as
$g(r)\rightarrow g(\beta)$.)

At this point one may ask how much of the LGT potential is simply due to the
modified Coulomb potential obtained in the previous section,
\be
V=a(-1/r+2T \log (r/\beta) )
\label{eq:32}
\ee
which results from the inclusion of ``reflections'' of the charge in the box.
Fixing the coupling constant at small r,
one obtains the curve plotted in figure \ref{fig:potpol} which
is not excluded by the LGT data.  Thus, the data of ref.\cite{MP87},
do not rule out the possibility that the ``spatial string tension'' is
actually very small.  However, at the lower temperature, $T \simeq 210$ MeV,
where the `wave functions' are being measured, the difference between the
modified Coulomb potential and the string potential are substantial (under the
assumption that the string tension remains the same). Therefore, by studying
the wave function one can discriminate between these two alternatives.  As we
will show in the following, the 'wave functions' for $\pi$ and $\rho$
given in ref.\cite{BDD91} seem to require the presence of a string potential
of the above magnitude.

With the  potential fixed, one can proceed further and solve the $d=2$
Schr\"odinger equation numerically.  In figure \ref{fig:Twave} we show
the corresponding wave functions for several temperatures for both
potentials eqn.\raf{eq:31} and eqn.\raf{eq:32}.  These results are
compared to data  at $T = 210$ MeV\footnote{The data are taken at $\beta =
6/{g^2} = 5.445$ with four time slices. Using the same $\beta$ and quark
mass but six time slices, the authors derive a temperature of $T_6 = \ 140$
MeV \cite{BOD91}.  This leads to a temperature of $T_4 = 210$ MeV for four
time slices.} These results
suggest that the confining effective potential, eqn.\raf{eq:31}, reproduces
the data well while the modified Coulomb potential is definitely too weak at
large distances\footnote{Provided the coupling constants for the logarithmic
and $1/r$ parts of the potential are the same.  Indeed, in this case, the
standard `asymptotic freedom' arguments do not apply.  Logarithmic
corrections are cut off at a scale of $1/T$, so both terms should have the
same coefficient, $\alpha_s(T)$.}.

We now consider the role of spin--dependent forces.  Comparing data for the
$\rho$ and $\pi$ channels (shown in figure \ref{fig:Twave}), one notes
significant differences.  The average size of the pion is significantly
smaller than that of the $\rho$.  At small distances ($r \leq 2\, a$), the
pion wave function is much larger\footnote{It is worth noting that the
$T=0$ pion wave function measured by Bernard {\em et al.} \cite{BDD91} is
in principle the same as the zero temperature Bethe--Salpeter
amplitude calculated by Chu, Lissia and Negele \cite{CLN91}.  Both drop
with an initial $e$--folding scale of $\sim 0.48$ fm which would give the
pion an rms--radius of $\sim 1/3$ fm.  This is substantially smaller than
the charge radius obtained in lattice gauge calculations of the
density--density correlation function \cite{Lea91}. In terms of phenomenology
\cite{BRW86}, the Bethe--Salpeter amplitude provides the ``intrinsic size''
of the pion (probing its $\bar{q} q$--component) while most of the charge
radius actually comes from the coupling of a virtual photon to the
$\rho$--meson ``cloud''.} with $|\psi(0)|^2$ being about four times greater
than in the $\rho$--meson case.

It is natural to ask whether this relative compactness of the pionic
$q \bar{q}$--wave function results chiefly from the perturbative spin--spin
interaction due to the exchange of high--frequency magnetic fluctuations.
The corresponding ($z$--averaged) interaction was discussed at the end of the
previous section, and it can simply be used after (i) switching from
standard QED to QCD units, $e^2 \rightarrow (4/3)(g^2 / 4 \pi)$, and (ii)
ignoring the annihilation contribution in the case of these isospin 1
channels.  Because the effects of this spin--spin interaction are not small,
we cannot simply estimate them using perturbation theory.  Rather, we must
re--solve our effective Schr\"odinger equation including this spin--dependent,
local interaction\footnote{In order to avoid difficulties arising from a
zero--range potential, we represent the $\delta$--function appearing in
$V_{Breit}$ by $\delta(x) = {\sin(\Lambda x)}/ \pi x$.  Here, $\Lambda$
represents a high frequency cutoff which we have chosen to be
$\Lambda = \pi / a$ in order to simulate the effects of a finite lattice
spacing, $a$.}.

   Two sets of results at temperatures of $210$ and $350$ MeV are shown in
figure \ref{fig:Swave}(a) and (b).  The three curves correspond to
$(S=1,S_z=\pm 1)$, $(S=1,S_z=0)$ and $(S=0,S_z=0)$ states.  A value of the
effective coupling constant of $\alpha_s=g^2/4\pi=  0.20$ was required
to describe the difference between $\pi$ and $\rho$ wave functions found in
LGT calculations.  This value of $\alpha_s$ is about 50\% larger than that
used in the central interaction, eqn.\raf{eq:31}.  This difference may have
several origins, but should not taken too seriously because, at
temperatures of $T \sim 210$ MeV, the wave functions are rather insensitive to
the strength of the central Coulomb potential\footnote{We point out that at
$T = T_c$ Karsch \cite{Kar88} gives a value of $\alpha_s(T_c) = 0.29$ for pure
glue. Also, the $T = 0$ value of $\alpha_s = 0.3$ extracted from
charmonium decay is of the same magnitude. \cite{CDD92}.}.  We remark that the
lattice wave functions are not normalized. In figure \ref{fig:Swave}(a)
the two different states of the $\rho$ have the correct relative
normalization, while the overall normalization of $\pi$ and
$\rho$--wave functions is arbitrary.  In figure \ref{fig:Swave}(b) all
three wave functions have the same normalization.

Our final point concerns the splitting of the screening masses.  From the
calculation displayed in figure \ref{fig:Swave}, we obtain
\be
E_\rho-E_\pi = 340 \,  {\rm MeV}
\ee
which can be compared with the results of direct measurements on the lattice
with six time slices ({\em i.e.,} at the lower temperature ($T= 140$ MeV),
which gives $\sim 440$ MeV.\cite{BOD91}

These predictions can be tested much more accurately by looking at the
difference between the longitudinally and transversely ($S_z = \pm 1, 0$)
polarized $\rho$ or by going to higher temperatures where our $d=2$
approximations should be better justified.  Although
the differences between wave functions for the different states of
the $\rho$ shown in figure \ref{fig:Swave} appears to be rather small for
both temperatures,  their ratio (particularly at short distances) should
provide sufficient information and should be less affected by systematic
errors.
The screening mass of these states differ by $\Delta M = 0.25 T$

The general question of the physical origin of spin--dependent splittings
in hadronic physics has been much debated in the
literature. Together with the Breit--type perturbative interactions,
interactions due to instantons, as proposed in refs.
\cite{Shuryak_A4,Rosner_Shuryak}, also produce a quasi--local interaction
of the form $(\vec\sigma_1 \cdot \vec\sigma_2)(t^a_1 \cdot t^a_2)$.
 One way to distinguish them is to consider their temperature dependence,
because the instanton density is expected to decrease rapidly at high T,
while the perturbative mechanism can even be enhanced as the bound states
discussed above become more compact.\\

Finally we mention that the effect of the Coulomb piece of the
interaction, eqn.\raf{eq:31}, on the wave function is tiny at those
temperatures where the lattice data have been obtained. Therefore, it seems
that these wave functions {\it require} the presence of a string--like
confining potential.
\newpage
\centerline{\bf 4. Conclusion and Discussion}
\ \\
Before summarizing our results, let us discuss the physical
meaning of the spatial correlation functions. It has been already
mentioned instead of yielding the more familiar spectrum of excitation
energies, spatial correlators provide only the spectrum of momenta
(or screening masses).  In distinction to the situation at zero temperature,
where these spectra are identical, the energy and momentum spectra
are by no means similar at finite temperature\footnote{We mention one more
simple example which emphasizes the difference between the two. Consider
non--interacting, massless fermions at high $T$. The excitation energy
spectrum obviously starts from zero and corresponds to the cut while the
momentum spectrum studied is a sequence of poles at $(2n+1)\pi T$.  The lowest
pole is our ``quark effective mass'' in the `funny space'.}.  Unlike the
former, momentum spectra display a very smooth $T$--dependence.
There are two reasons for this fact.  First, the quark effective  mass
consists of two parts, $m_{\rm eff} = \sqrt{\pi^2 T^2 +
m_{dyn}^2} $.  The first increases with $T$ and the second decreases as $T
\rightarrow T_c$ with the result that the total value is roughly constant
below $T_c$.  Second, due to some dynamical reason (as yet unclear), the
effective potential for spatial propagation is also temperature
independent\footnote{Apparently, this potential ``does not notice" even a
very strong first--order deconfinement transition which is observed in pure
gluonic theory.  Notice that this potential has not yet been
evaluated when dynamical quarks are present.}.  As a result, there is a
correspondence between the poles (and the corresponding ``wave functions")
for the $T=0$ and high--$T$ cases.  In this (but only in this) sense,
hadronic degrees of freedom smoothly traverse this boundary.

It is also worth recalling that the trend seen in the momentum
spectra discussed above is quite different from what is known about the trends
of the energy spectra.  Various theoretical models  suggest that the
energy gaps (usually still referred to as ``hadronic masses") decrease
with $T$.  (See, {\em e.g.,} recent  calculations based on hadronic
scattering data \cite{Shuryak_pot}.) There are suggestions that they
even vanish at $T=T_c$ \cite{BR91} so that, at higher temperatures,
the energy spectrum is just a cut containing excitations
of any energies starting from zero.  There is no contradiction between these
two pictures: Propagation in space and in time at high $T$ are no longer
simply related.  Each can tell us something interesting about
the properties of the QGP.

In this paper we have addressed not only these momentum (or ``screening mass")
spectra but also the corresponding wave functions.  We have shown that the
spatial correlation functions at high temperatures can be quite well
understood in a `funny space' obtained by interchanging time and space axes.
The two main ingredients of our calculations are the
quark effective mass, $m_{eff}=\pi T$, and the effective potential, $ V_{eff}$,
which is a combination of a Coulomb and a confining part and is supplemented by
a short--range spin--spin interaction.  We find these ingredients to provide
a good description of the wave functions measured in \cite{BDD91}.  We further
find that, at the temperatures for which lattice wave functions have been
measured ($T \simeq 210 \rm MeV$), the effect of the Coulomb piece of the
effective
interaction is small. Hence, we conclude that the lattice data require a
relatively strong potential --- much stronger than the modified Coulomb
potential which would arise from dimensional reduction arguments \cite{HZ91}.
Thus, as expected, dimensional reduction is not valid at temperatures as low
as $T \simeq 210$ MeV.  Additional lattice calculations at considerably higher
temperatures would be needed in order to check the validity of dimensional
reduction.

The issue certainly can be pursued further both numerically and analytically.
In particular, better data and data at higher temperatures are needed
for further tests of the theory, as well as more detailed studies
of the effective potential by itself {\it a la} \cite{MP87}.
For example, calculations of the effective potential with dynamical quarks are
needed. Obviously, further attempts can be made in order to get a smooth
matching between the traditional potential model description of hadrons at
$T=0$ and our approach at high $T$.  This could be done by solving
the complete three dimensional Dirac equation
with anti-periodic  boundary conditions. Splittings between different
spin--isospin components can be studied in greater detail.

What have we actually learned from these ``wave functions" in simple physical
terms?   Consider a spatial correlation function at distance, $z$,
with the two ends taken for simplicity at the same instant of time, and let
us describe qualitatively the typical paths contributing to it.  (This
language is convenient because it is the same in physical and `funny' space.)
At $T=0$ the light quarks may travel by any path\footnote{
As usual, one can say that the correlation is caused by  pair production
somewhere in the middle, and its existence does not contradict causality
because one cannot use it for signal transfer.} deviating from the
$z$ axis by $\delta t \sim \delta x \sim \delta y \sim z$ (see figure
\ref{fig:path}(a)) which leads to power--like correlation functions. At high
$T$, for non--interacting quarks, the deviations in time are smaller,
$\delta t \sim 1/T$, due to the cancellation effects described in the
introduction, and the correlator is $\exp(-Mz)$.   As a result,
$\delta x \sim \delta  y \sim \sqrt{z / T}$.  (See figure \ref{fig:path}(b).)
These are the paths of ``heavy" non--interacting quarks in the `funny space'.
Finally, for high $T$ and interacting quarks, quark and
antiquark now travel along paths which are correlated within a
distance governed by the ``wave functions" considered above.  (See figure
\ref{fig:path}(c).)  Their deviation from the straight line remains as in case
(b).

The perturbative Coulomb interaction in `funny space' arises
physically from the perturbative magnetic interaction
of two color currents in original physical space. In the very high $T$ limit,
when this effect dominates, the distance between them is
$\delta x \sim \delta  \sim 1/(gT)$.  Note that in a quark--gluon plasma the
magnetic interaction is screened (if at all) only on a parametrically larger
scale of $R_m \sim 1/(g^2 T)$.  Thus, there is no contradiction here.

Of course, the origin of the non--perturbative ``confining" potential remains
unclear.  It is, in any event, due to some non--perturbative $G_{xy}$ field
which is the magnetic field both in physical and `funny' spaces.

It is important to distinguish between magnetic and electric interactions
which are very different even in the more familiar electromagnetic plasma.
The electric fields are screened and appear only as high frequency
fluctuations.  Magnetic fields are not, and they
can produce  very spectacular structures (as in the sun). In QCD the
additional non--trivial feature is that the long--range magnetic fields are
self--interacting, so their structure at high T is not understood at all.

As a final remark, let us try to answer the following question: Are
the phenomena considered here subject to experimental observation?
Correlations of electromagnetic currents are the source of photon and
dilepton production from the plasma. Therefore,
correlators similar to those introduced here will be essential in determining
the effects of perturbative and non--perturbative gluomagnetic fields
on these reactions. Such correlators are not easily obtained from
LTG calculations. Here we have shown that a simple dynamical model is capable
of reproducing the results of these more complicated LGT calculations for zero
frequency correlators. This calculations have established a consistency between
the zero frequency correlator and the spatial potential of ref.\cite{MP87}.
Moreover, they reveal the absolute necessity of a linearly rising piece in this
potential. Given the success of this dynamical model and its relative ease of
implementation, it seems both practical and promising to use it for more
detailed calculations of photon and dilepton production.
Thus, although much additional
work needs to be done, it may turn out that the topic of this paper is
not as academic as it may appear at first sight.
\ \\
\vskip 1cm

\centerline{\bf Acknowledgements}

\vskip 1cm

We are particularly grateful to Carleton DeTar for communications and helpful
advice. He also provided us with the data file for the lattice wave functions.
This work was initiated by discussions with Tom DeGrand at Quark--Matter--91
where he reported the data under consideration.  Discussions with I.~Zahed
at the initial stage of this work was also very important. Finally,
we want to acknowledge fruitful discussions with H.A.~Bethe on the spin--spin
interaction.
\newpage

\centerline{{\bf Appendix}}
\ \\
In this appendix we will describe the details of the Schr\"odinger equation for
an $e^+ e^-$ pair.
The wave function is characterized by the Matsubara frequency of the pair
in `funny space', $p_z = 2 n \pi T$.  The frequencies of the electron and
positron must sum to this value but their difference is free.  The wave
function for the $e^+ e^-$ pair can thus written as
\be
\Psi^n = \sum_{k=-\infty}^{\infty} c_k \psi^n_k
\ee
where $n$ is the sum of the Matsubara frequencies and $-\infty \leq k \leq
+\infty$ their difference.
The coefficients, $c_k$, are determined from the solution of the
following Schr\"odinger equation:
\be
\left( \pi T ( |n+k+1| + |n-k-1|) +
        \frac{p_{\bot,1}^2}{2 (n + k +1) \pi T} +\right.&&
\nonumber \\
\left.
\frac{p_{\bot,2}^2}{2(n - k - 1) \pi T} + V_0 \right) \psi^n_k
 + \sum_{l=+1}^{+\infty} V_{2 l} [\psi^n_{k-l} + \psi^n_{k+l} ]
&=& E^n \psi_k^n
\label{cpl}
\ee
where $V_o$ is the effective Coulomb Potential (eqn.\raf{eq:2.3})
and
\be
V_{2 l}(\vec{r}_\bot) =
\frac{T}{(2 \pi)^2} \int d^2 k_\bot e^{i \vec{r}_\bot \cdot \vec{k}_\bot}
\frac{1}{k_\bot^2 + ( 2 l \pi T)^2}
\ee
are the higher Matsubara components of eqn.\raf{eq:2.2}.

These energies and wave functions are the immediate building blocks for the
various correlators which we can construct.
In the original, unrotated space, these are given by
\be
C(\vec{r}_\bot,z,\tau) &=&
\int d^2x_\bot <\hat{T}[\hat{\cal O}(\vec{x}_\bot,z = 0,\tau=0)
\bar{q}(\vec{x}_\bot,z,\tau) \Gamma q(\vec{x}_\bot + \vec{r}_\bot,z,\tau)]>
\nonumber \\
& = & \sum_n \exp(-i \omega_n \tau) C(\vec{r}_\bot,z,\omega_n)
\label{corr}
\ee
where the ``source" $\hat{\cal O}$ is an operator
which creates quark and antiquark at $z,\tau = 0$,
$\Gamma$ the appropriate Dirac matrix for the meson under
consideration and $\omega_n$ are the usual Matsubara frequencies. The Matsubara
components of the above defined correlation functions are then related to the
solution of the Schr\"odinger equation \raf{cpl} by
\be
C(\vec{r}_\bot,z,\omega_n) = \sum_\alpha  a_\alpha
\exp(-M_\alpha^n z) \Psi_\alpha^n (\vec{r}_\bot)
\ee
The index, $\alpha$, labels the eigenstates of eqn.\raf{cpl},
$M_\alpha^n$ are the
associated screening masses of these states which are identical with the
eigenvalues of the Schr\"odinger equation \raf{cpl} and $a_\alpha$ are
some coefficients which we do not further specify.
In the limit of large $z$ only the ground state survives and we have
\be
\lim_{z \rightarrow \infty}
C(\vec{r}_\bot,z,\omega_n) \propto \exp(-M_0^n z) \Psi_0^n (\vec{r}_\bot)
\ee
where $\Psi_0^n$ denotes the solution of eqn.\raf{cpl} of lowest energy for
given z--momentum in `funny space' $p_z = 2 \pi n T = \omega_n$.

The primary focus of LGT so far has been the zero
frequency correlator ({\em i.e.,} $n = 0$). This is achieved by integrating
the correlator, eqn.\raf{corr}, over $\tau$. In addition, an uncorrelated
source operator of the form
\be
{\cal O}_{uncorr.} (\vec{x}_\bot,z,\tau) = \int d \tau' \, d^2 x'_\bot
\, \bar{q}(\vec{x}_\bot,z,\tau) \, \Gamma \, q(\vec{x}_\bot + \vec{x}'_\bot,z,
\tau+\tau')
\ee
is used.  This source sends off a quark--antiquark pair with vanishing relative
Matsubara frequency and transverse momentum.

Finally, notice that transverse wave functions result from correlations
between quark and antiquark at equal time because both operators are defined
at the same  time, $\tau$, in eqn.\raf{corr}.

\newpage

\centerline{\bf Figure captions:}
\ \\
\bfg
\caption{Schematic picture of replications leading to the two--dimensional
Coulomb potential \protect\raf{eq:2.3} }
\label{fig:2dcoul}
\efg

\bfg
\caption{Wave function (full line) and potential (dashed line)
for two--dimensional positronium with $x = r (eT)$}
\label{fig:pot}
\efg

\bfg
\caption{Comparison of finite temperature Coulomb potentials. The solid lines
correspond to the potential \protect\raf{eq:2.3}, while the dashed ones
correspond to the two--dimensional potential, $2 T \protect\log (r T)$.}
\label{fig:Potcomp}
\efg

\bfg
\caption{The potential of ref.\protect\cite{MP87}.
The finite temperature ($T > T_c$) space--like potential corresponds to the
full line, the zero temperature time-like potential to the dashed line.
The dashed-dotted line
 represents the
finite temperature ($T > T_c$) space-like potential and the
short-dashed--long-dashed line the
modified Coulomb potential according to eqn.\protect\raf{eq:2.3}.
}
\label{fig:potpol}
\efg

\bfg
\caption{Wave function for the $\rho$-meson as function of r/a
($a = 0.22 \,\rm fm$) for
different temperatures. The upper set of
curves corresponds to wave functions obtained with
the modified Coulomb potential (eqn.\protect\raf{eq:32})
the lower set with the confining potential (eqn.\protect\raf{eq:31}).
Temperatures: 150~MeV (full line), 250~MeV (short-dashed line) and
350~MeV (long-dashed line). Data are form ref. \protect\cite{BDD91}
and correspond to $T \simeq 210 \,\rm MeV$.}
\label{fig:Twave}
\efg

\bfg
\caption{Wave functions for  $\rho$ and $\pi$ at $T= 210 \, \rm MeV$ (a) and
$T = 350 \, \rm MeV$ (b) as a function of $r/a$ ($a = 0.23 \, \rm fm$).
The full line and the short--dashed line correspond to
the $\rho$ with spin quantum number $S_z = \pm 1$ and $S_z = 0$ respectively.
The long--dashed line represents the pion wave function. In (a) arbitrary
normalization was used while in (b) all wave functions have the same
normalization. Data are form ref.\protect\cite{BDD91}. }
\label{fig:Swave}
\efg

\bfg
\caption{Different paths contributing to a equal time correlator of two
currents separated by a distance $z$. In the $T=0$ case (a) the transverse
deviation of the paths is $r_\perp \sim z$ while at high $T$ for
{\it non-interacting}
quarks (b) it is smaller, $r_\perp \sim \protect\sqrt{z/T}$. The
{\it bound}
quarks (c) still deviate from the axis by $r_\perp \sim \protect\sqrt{z/T}$,
but
their relative distance is defined by the size of the bound state $r_{bound}$.}
\label{fig:path}
\efg

\end{document}